\documentclass[twoside]{dis09}
\usepackage[latin1]{inputenc}
\usepackage[dvips]{graphicx,epsfig,color}
\usepackage{wrapfig,rotating}
\usepackage{amssymb,amsmath,array}

\begin{document} 
\title{Jet cross sections and $\alpha_s$ in deep inelastic scattering and photoproduction at HERA}

\author{Claire Gwenlan\footnote{On behalf of the ZEUS Collaboration \cite{slides}.} 
%
%
\vspace{.3cm}\\
%
Department of Physics, University of Oxford, \\
Denys Wilkinson Building, Keble Road, Oxford. OX1 3RH. UK.\\
email: c.gwenlan1@physics.ox.ac.uk
%
}

\maketitle

\begin{abstract}
Recent ZEUS measurements of inclusive-jet and dijet cross sections in neutral current 
deep inelastic $ep$ scattering at HERA are presented. 
The data correspond to more than a two-fold increase in statistics compared to previous studies. 
The cross sections are measured in the Breit frame, 
for boson virtualities of $Q^2>125$ ${\rm GeV}^2$, as functions of various kinematic 
and jet observables.  
The data are found to be well described by NLO QCD and have the potential to 
constrain the gluon density in the proton.
Two new extractions of the strong coupling, $\alpha_s$, are also presented:  
the first is determined from the inclusive-jet neutral current DIS measurement presented here,   
while the second is from a re-analysis of previously published 
data on inclusive jet photoproduction. 
Both measurements are of competitive precision and in agreement with the world average.
\end{abstract}

\section{Introduction}
Jet production in $ep$ collisions at HERA provides an important 
testing ground for perturbative Quantum Chromodynamics (QCD), 
as well as giving direct access to the strong coupling constant, 
$\alpha_s$. Furthermore, HERA jet measurements are sensitive 
to the parton distribution functions (PDFs) of the 
proton. 

At HERA, there are broadly two kinematic regimes: 
Deep Inelastic Scattering (DIS), where the virtuality of the 
exchanged boson is large ($Q^2 \gg 1$ ${\rm GeV}^2$), and photoproduction ($\gamma p$), 
which proceeds via the exchange of a quasi-real photon ($Q^2 \approx 0$).
In this contribution, recent measurements of inclusive-jet 
\cite{zeusprel-09-006} and dijet \cite{zeusprel-07-005} 
production in neutral current (NC) 
DIS are presented. The data used in these measurements  
correspond to more than a two-fold increase in statistics compared to 
previous studies. 
The measurements are sensitive to the parton content of the proton, 
especially the gluon density at high momentum fractions, and 
may serve as valuable inputs to future QCD fits for the PDFs.
In addition to the jet cross section measurements, two new determinations  
of $\alpha_s$ are presented: the first is extracted from 
the NC DIS inclusive-jet measurement \cite{zeusprel-09-006} presented here, and the second  
is derived from a re-analysis \cite{zeusprel-08-008} of 
previously published data \cite{plb:560:7} on inclusive jets in $\gamma p$.

\section{Data selection and correction}
The data used for the jet cross section measurements were collected 
with the ZEUS detector at HERA. 
For the inclusive-jet analysis, the data are from the $05-06$  
running period and correspond to $188 ~{\rm pb}^{-1}$. 
For the dijet case, the data are from $98-00$ and $04-05$ 
running, combined, corresponding to a total of $209 ~{\rm pb}^{-1}$. 
During these times, HERA operated with protons of energy 
$E_p=320$ GeV and  electrons of energy $E_e=27.5$ GeV.

The phase space of the measurements is restricted to the region 
$Q^2 > 125$ ${\rm GeV}^2$ and $-0.65 < \cos(\gamma_{had}) < 0.65$, where 
$\gamma_{had}$ is the polar angle of the hadronic 
system\footnote{In Quark Parton Model type events, $\gamma_{had}$ 
is the angle of the scattered quark.}. For the dijet case, 
a further requirement of $Q^2 < 5000$ ${\rm GeV}^2$ was imposed: this 
restricts the data to a region where the contribution 
from $Z^0$ exchange is negligible.

Jets were reconstructed in the Breit\footnote{In NC DIS, 
the Breit frame is preferred to conduct the jet search 
since jet production is directly sensitive to hard QCD sub-processes at 
$\mathcal{O}(\alpha_s)$ (and higher); in the Born level process ($eq\rightarrow eq$), 
the virtual boson $V^*$ (where $V^*=\gamma,Z^0$) is absorbed by 
the struck quark, which is back-scattered with zero transverse momentum 
with respect to the $V^*$ direction. At leading order in $\alpha_s$, 
the Boson-Gluon-Fusion ($V^* g \rightarrow q\bar{q}$) and QCD Compton 
($V^* q \rightarrow qg$) processes give rise to 
two hard jets with opposite transverse momenta.} frame \cite{breit}, using the 
longitudinally invariant $k_T$-clustering algorithm \cite{np:b406:187} 
in the inclusive mode \cite{pr:d48:3160}, 
and were required to lie in the pseudo-rapidity range $-2 < \eta_{\rm Breit} < 1.5$. 
For the inclusive-jet analysis, events were selected if they 
contained at least one jet with $E_{T,Breit} > 8$ GeV. For the 
dijet case, at least two jets were required with 
$E_{T,Breit}^{1,2} > 12, 8$ GeV: the asymmetric cut 
is made in order to avoid infra-red sensitive regions 
where the next-to-leading order (NLO) QCD programs are not reliable.

The data were corrected for detector efficiency and 
acceptance effects using the ARIADNE \cite{ariadne} and 
LEPTO \cite{lepto} Monte Carlo (MC) models. The MC programs 
were also used to correct the measured cross sections for 
QED radiative effects and the running of $\alpha_{em}$.

The typical statistical precision on the final cross section 
measurements is $\sim 1-5\%$. The dominant experimental systematic 
comes from the uncertainty on the jet energy scale (known to 
$\pm 3\%$ for $E_{T,Lab}<10$ GeV and $\pm 1\%$ for higher jet-transverse-energies), 
leading to effects on the cross sections of typically $5-10\%$. 
The second-most-important is the model uncertainty arising 
from using ARIADNE versus LEPTO to correct for detector effects 
(this leads to uncertainties of typically $\lesssim 3\%$ on the cross sections). 

\section{NLO QCD calculations}
\label{sec:nlo}
The NLO QCD ($\mathcal{O}(\alpha_s^2)$) calculations used to compare with the data 
were obtained using the program DISENT \cite{disent}. 
The factorisation scale was  taken to be $\mu_F=Q$ and the 
renormalisation scale\footnote{Other choices of scale were also checked.}   
was taken to be $\mu_R=E_{T,Breit}$ (of each jet) for the 
inclusive-jet measurement and $\mu_R=(Q^2+\overline{E}^2_{T,Breit})^{1/2}$ for the dijet case, 
where $\overline{E}_{T,Breit}$ is the average transverse energy of the two jets.
The strong coupling was calculated at two loops using 
$\Lambda^{(5)}_{\rm{\overline{MS}}} = 226$ MeV,
corresponding to $\alpha_s(M_Z)=0.118$. The calculations were performed 
using the ZEUS-S \cite{pr:d67:012007} proton PDF for the inclusive-jet   
and CTEQ6 \cite{cteq6} for the dijet measurement. 
In order to compare directly with the data, the NLO QCD 
predictions were corrected for hadronisation using the average of the 
corrections obtained from ARIADNE and LEPTO. 
The inclusive-jet predictions were also corrected for $Z^0$ exchange. 

Several sources of theoretical uncertainty were considered: uncertainties   
due to neglected terms beyond NLO in the perturbative expansion; 
uncertainties from the input proton PDFs; uncertainties from $\alpha_s$; 
hadronisation correction uncertainties; and those  
due to the choice of factorisation scale. 
The dominant theoretical uncertainty is due to the contribution from 
terms beyond NLO, which was estimated by varying the renormalisation 
scale by the (conventional) factors of $\frac{1}{2}$ and $2$. This resulted in 
variations in the predicted cross section of $\sim 5-20\%$. 
All other sources typically resulted in only small changes to the predicted cross sections.

\section{Results}

\subsection{Jet cross sections}
\begin{figure}[Htp]
\begin{minipage}{6.8cm}
\hspace{-0.5cm}\includegraphics[width=8.5cm,height=8.5cm]{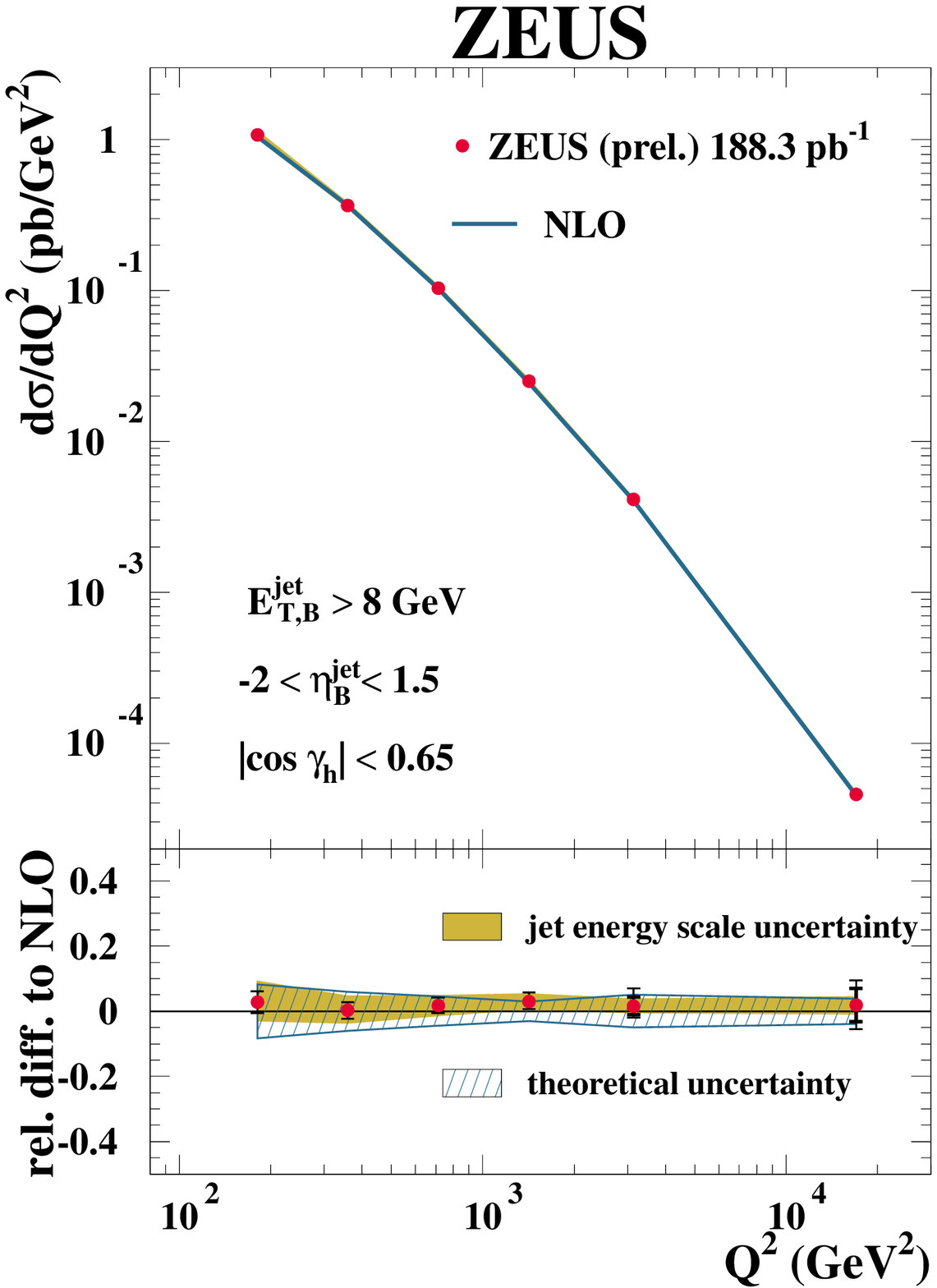}
\caption{Single differential cross section $d\sigma/dQ^2$ for inclusive-jet 
production with $E_{T,Breit}>8$ GeV and $-2<\eta_{Breit}< 1.5$ 
from ZEUS~\cite{zeusprel-09-006}.\label{fig:inclusive}}
\end{minipage}
\hspace{0.4cm}
\begin{minipage}{6.8cm}
\vspace{-0.45cm}
\hspace{-0.4cm}
\includegraphics[width=7.cm,height=9cm]{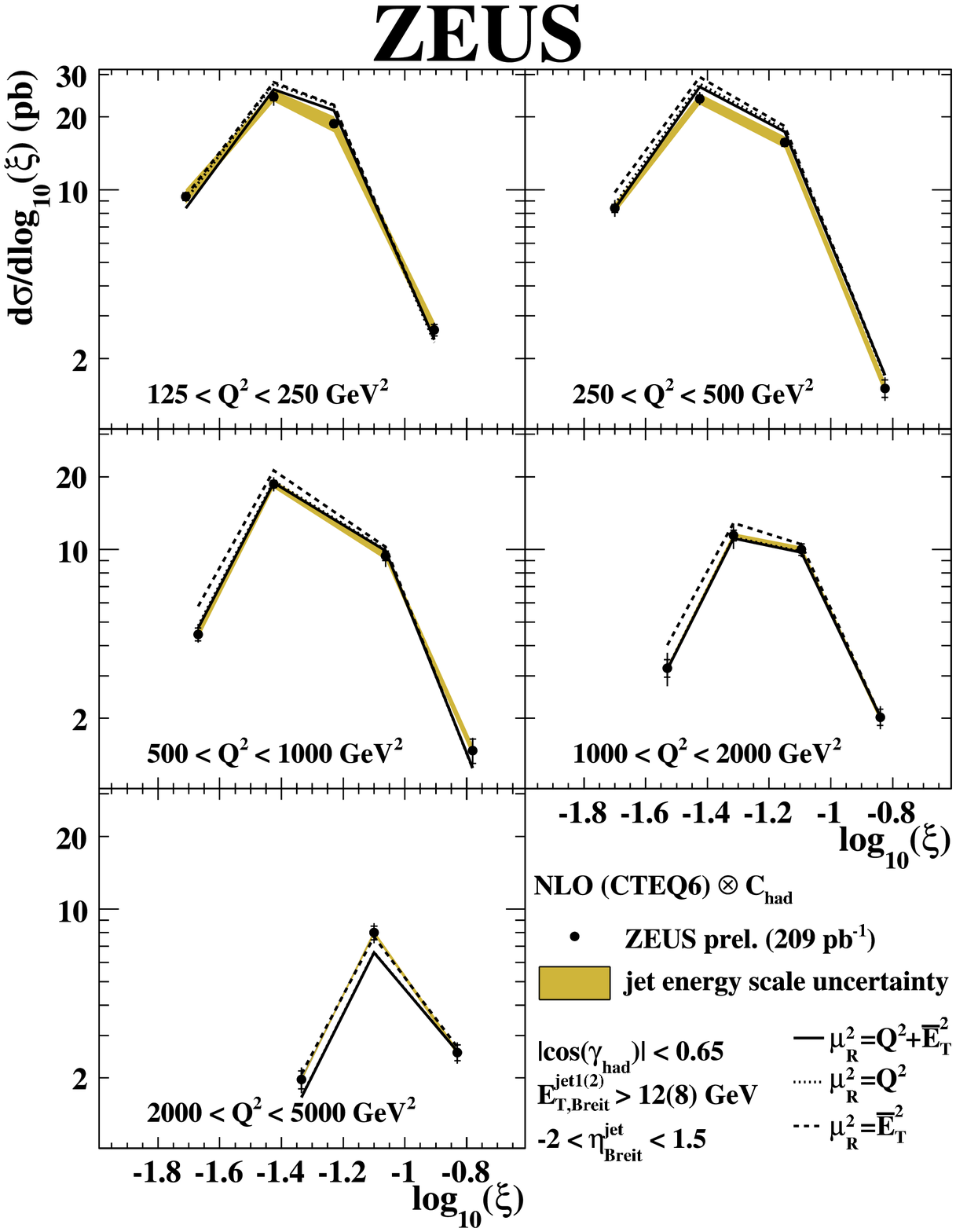}
\vspace{-0.5cm}
\caption{Double differential dijet cross sections as a function of 
$\log_{10}(\xi)$ in regions of $Q^2$ from ZEUS~\cite{zeusprel-07-005}.\label{fig:dijet}}
\end{minipage}
\end{figure}

The inclusive-jet cross sections were measured differentially as functions of $Q^2$, 
$E_{T,Breit}$ and $\eta_{Breit}$. 
Figure~\ref{fig:inclusive} shows the single differential 
cross section as a function of $Q^2$: 
the inner and outer error bars show the statistical and uncorrelated systematic   
uncertainties respectively; the shaded band indicates the bin-by-bin correlated 
systematic uncertainty due to the jet energy scale; 
and the hashed area gives the theoretical uncertainty. 
The results show that the data are well described by NLO QCD over the measured range. 
Except at the very highest $Q^2$ values, the theoretical uncertainties  
dominate over the experimental. 
Double differential cross sections as a function of $E_{T,Breit}$, in 
$6$ bins of $Q^2$, were also measured. Previously, a similar 
measurement of inclusive jets in NC DIS \cite{pl:b547:164} was included\footnote{Along with a measurement 
of dijets in photoproduction \cite{epj:c23:615}.} in a NLO QCD fit 
\cite{epj:c42:1} performed by the ZEUS collaboration. The results 
showed that the HERA jet data were able to significantly constrain the gluon PDF at 
mid-to-high values of the proton momentum fraction. The new measurement 
presented here is in the same region of phase space but corresponds 
to $\sim 4.5$ times the statistics. The very high experimental precision  
 makes it a promising candidate for 
inclusion in future QCD analyses, potentially providing substantial 
further constraints on the gluon density in the proton.

The dijet cross sections were measured differentially as functions of a number of kinematic 
and jet observables, including $Q^2$, $M_{jj}$ (the dijet invariant mass), 
$\overline{E}_{T,Breit}$ and $\log_{10}(\xi)$\footnote{$\xi = x_{Bj} \left 
( 1 + M_{jj}^2/Q^2\right )$ is the parton momentum fraction, where $x_{Bj}$ is the Bjorken-$x$ variable.}. 
The data are generally well described by the predictions of NLO QCD.
Double differential cross sections as a 
function of $\log_{10}(\xi)$, in $5$ regions of $Q^2$, were also measured 
(Fig.~\ref{fig:dijet}). 
The data are reasonably well described by NLO QCD, although there is 
some sensitivity to the choice of renormalisation scale. At large $Q^2$, 
the statistical uncertainties dominate. The contribution 
of gluon-induced events to the total cross section was also estimated as a function of $Q^2$ and $\xi$. 
It is observed that the contribution from gluon-induced processes is at least $30\%$, 
even for high values of $Q^2$ ($\approx 3000$ ${\rm GeV}^2$) and $\xi$.
The uncertainties on the predicted cross sections, arising from the 
gluon PDF, are found to be substantial. 
Therefore, these data are also expected to provide potentially 
strong constraints on the gluon content of the proton.

\subsection{Strong coupling, $\alpha_s$}
An NLO QCD analysis was performed on the single-differential NC DIS inclusive-jet cross section 
as a function of $Q^2$ in order to determine a value of $\alpha_s(M_Z)$ (see Fig.~\ref{fig:inclusive}). 
Only the region $Q^2>500$ ${\rm GeV}^2$ was used, giving the best overall uncertainty 
on the extracted value.   
The QCD calculations were performed using the program DISENT 
(as described in Sec.~\ref{sec:nlo}), using $5$ different sets of PDFs 
from the ZEUS-S fit, each with different values of $\alpha_s(M_Z)$. The value 
of $\alpha_s(M_Z)$ used in each cross section calculation was that associated with 
the PDF used. The $\alpha_s$ dependence of the predicted cross sections, 
in each $Q^2$ bin, was parameterised using a function quadratic in $\alpha_s(M_Z)$. 
The value obtained is: 
$\alpha_s(M_Z) = 0.1192 \pm 0.0009 {\rm (stat.)}^{+0.0035}_{-0.0032}
{\rm (exp.)}^{+0.0020}_{-0.0021}{\rm (th.)}$, corresponding to a total uncertainty of $3.7\%$. 
The experimental uncertainties are dominated by the contribution from the jet energy scale ($\pm 1.9\%$) 
while the theoretical uncertainties are dominated by terms beyond NLO ($\pm 1.8\%$), 
as evaluated using the method of Jones {et al.} \cite{jones}. 
The uncertainties from other sources (proton PDFs, 
hadronisation corrections and $\mu_F$) are generally small. 

A similar method was employed to extract a new value of $\alpha_s(M_Z)$ from a previously  
published measurement of inclusive-jets in $\gamma p$ \cite{plb:560:7}. 
The single differential inclusive-jet cross section as a function of 
$E_T^{jet}$ was used for the determination.
The NLO QCD calculations were performed using the program of 
Klasen, Kleinwort and Kramer \cite{epj:c1:1}. The factorisation and renormalisation scales 
were taken to be $\mu_R=\mu_F=E_T^{jet}$ and the photon PDF was GRV-HO \cite{pr:d45:3986}. 
With respect to the original publication, the input proton PDF was   
updated to MRST01 \cite{epj:c23:73} (c.f. MRST99) and the method of Jones {et al.} \cite{jones} was used to evaluate the 
uncertainties from terms beyond NLO. The value of $\alpha_s(M_Z)$ extracted is: 
$\alpha_s(M_Z) = 0.1223 \pm 0.0001 {\rm (stat.)}^{+0.0023}_{-0.0021}
{\rm (exp.)}^{+0.0029}_{-0.0030}{\rm (th.)}$, corresponding to a $3.1\%$ total uncertainty. 
The central value is almost identical to that obtained in the original publication using the same data. 
The dominant experimental uncertainty arises from the jet energy scale  
($\pm 1.5\%$) and the theoretical 
uncertainty is again dominated by terms beyond NLO ($\pm 2.4\%$)\footnote{The method of Jones {et al.} 
results in a smaller theoretical uncertainty than in previously.}.

Both measurements are in agreement with other determinations from HERA, 
and with the world average, and are of competitive precision. NNLO calculations 
will be needed to further improve the theoretical uncertainties in the future.

\section{Summary}
Inclusive-jet and dijet cross sections have been measured in NC 
DIS at HERA, for $Q^2>125$ ${\rm GeV}^2$, using more than a 
two-fold increase in statistics compared to previous studies. 
The measurements are well described by the predictions of NLO QCD 
and are sensitive to the proton PDFs, especially    
the gluon density at high momentum fractions.  
The measurements are therefore natural 
candidates for inclusion in future QCD fits for the proton PDFs. 

In addition, a value of  $\alpha_s(M_Z)$ has been extracted from the  
NC DIS inclusive-jet single-differential cross section as a function 
of $Q^2$ (for $Q^2>500$ ${\rm GeV}^2$). 
A further determination has been obtained from a re-analysis of 
previously published data on inclusive-jet $\gamma p$. Both extractions are of competitive 
precision and in agreement with the world average.


\begin{footnotesize}



%

\end{footnotesize}


\end{document}